\documentclass[journal,10pt]{IEEEtran}

\usepackage{amsfonts}
\usepackage{amssymb}
\usepackage{stfloats}
\usepackage{cite}
\usepackage{graphicx}
\usepackage{psfrag}
\usepackage{subfigure}
\usepackage{amsmath}
\usepackage{array}
\usepackage{algorithm}
\usepackage{algpseudocode}
\usepackage[affil-it]{authblk}
\usepackage[english]{babel}
\usepackage{blindtext}
\usepackage{url}
\usepackage{epstopdf}
\usepackage{flushend}
\usepackage{verbatim}
\usepackage{bm}
\usepackage{multirow}
\usepackage{booktabs}
\usepackage{caption}

\newtheorem{Prob}{Problem}

\title{Fingerprint-based Localization using Commercial LTE Signals: A Field-Trial Study}

\author{Heng Zhang$^{\dagger}$, Zhichao Zhang$^{\dagger}$, Shunqing Zhang$^{\dagger}$, Shugong Xu$^{\dagger}$ and Shan Cao$^{\dagger}$\\
$^{\dagger}$ Shanghai Institute for Advanced Communication and Data Science, \\
Key laboratory of Specialty Fiber Optics and Optical Access Networks, \\
Shanghai University, Shanghai, 200444, China\\
Email:\{hengzhang, zhichaozhang, shunqing, shugong, cshan\}@shu.edu.cn}

\begin{document}
\maketitle

\begin{abstract}
Wireless localization for mobile device has attracted more and more interests by increasing the demand for location based services. Fingerprint-based localization is promising, especially in non-Line-of-Sight (NLoS) or rich scattering environments, such as urban areas and indoor scenarios.
In this paper, we propose a novel fingerprint-based localization technique based on deep learning framework under commercial long term evolution (LTE) systems. Specifically, we develop a software defined user equipment to collect the real time channel state information (CSI) knowledge from LTE base stations and extract the intrinsic features among CSI observations. On top of that, we propose a time domain fusion approach to assemble multiple positioning estimations. Experimental results demonstrated that the proposed localization technique can significantly improve the localization accuracy and robustness, e.g. achieves Mean Distance Error (MDE) of 0.47 meters for indoor and of 19.9 meters for outdoor scenarios, respectively.
\end{abstract}


\section{Introduction} \label{sect:intro}
Localization has been identified as one of the most popular applications in the modern society, especially when massive amounts of devices are connected with each other \cite{del2017survey}. Various types of services, including autonomous driving \cite{geiger2012we} and indoor navigation \cite{Pecoraro2018CSI, shi2018accurate}, require high resolution localization information in outdoor and indoor environments, which motivates continuous research interests in recent years \cite{ye2017neural,shi2018accurate,Pecoraro2018CSI}. Although {\em Global Navigation Satellite Systems} (GNSSs), such as {\em Global Positioning System} (GPS), can provide continuous localization information in the outdoor environments, the performance degradation usually happens when the satellite signals are blocked, e.g. in some urban canyons and indoor environments\cite{Pecoraro2018CSI},
which triggers the effort to support cellular-based localization techniques.

Several cellular network positioning techniques have been considered in Release 9 of the 3rd Generation Partnership Project (3GPP) document such as Observed Time Difference Of Arrival (OTDOA) and Enhanced Cell ID (E-CID). Radio Frequency (RF) fingerprint methods are also included in LTE Release 9 due to no additional standardization and good resistance to NLoS environments\cite{r9}. RF fingerprint methods  contain two phases, where a fingerprint map was generated offline during the training phase, and an online location estimation algorithm was performed based on real time fingerprint measurements.
The Received Signal Strength Indicator (RSSI) or Reference Singal Received Power (RSRP) messages are usually considered as location fingerprint due to its easy access to the terminals.
However, the RSSI/RSRP based localization often suffers from wireless fluctuation due to the potential shadowing or signal blockage effects, and the overall localization accuracy is limited \cite{wu2017mitigating}. To overcome this limitation, the CSI has been proposed as a suitable candidate for fingerprint information \cite{wang2017csi} , especially in the wide band system such as WiFi. Another critical condition for the practical deployment of localization schemes is the fast online matching algorithm, where K-nearest neighbor (KNN) \cite{Pecoraro2018CSI},  weighted K-nearest neighbor (WKNN) \cite{machaj2011rank} and support vector machine (SVM) \cite{lin2014group} are commonly adopted. Although the above unsupervised learning approaches provide reasonable online matching delay, the localization accuracy as well as the robustness performance is in general limited \cite{wang2017csi}.


In this paper, to address the above issues, we propose a novel fingerprint-based localization technique based on deep learning framework under commercial LTE systems. Specifically, we develop a software defined user equipment to collect the real time CSI knowledge from LTE base stations and extract the intrinsic features among CSI observations. On top of that, we propose a time domain fusion approach to assemble multiple positioning estimations. Experimental results show that the proposed approach can significantly improve the localization accuracy and robustness if compared with traditional online positioning algorithms. The main contributions of this paper are summarized below.

\begin{itemize}
\item{\bf Improved Accuracy with Deep Learning} Since the traditional matching algorithm is difficult to represent the mapping relation between the high-dimensional CSI fingerprint and the instantaneous terminal position, we approximate this relation through deep neural networks (DNN). With some field trial measurement results, the localization accuracy MDE can be improved quite significantly, e.g. from 1.31m to 0.47m and from 90.9m to 19.9m for indoor and outdoor environments, respectively, compared with KNN.
\item{\bf Robust Design with Temporal Fusion} As the time-varying nature of CSI, we also propose a time domain fusion approach that can capture the temporal correlation of different CSI samples. By eliminating the microscopic fluctuations of CSI samples, the extracted temporal correlation information can greatly improve the robustness performance in terms of the maximum distance error, e.g. from 3.19m to 1.15m and 297m to 64.8m for the indoor and outdoor scenarios, respectively.
\end{itemize}

The rest of the paper is organized as follows. In Section~\ref{sect:sys}, we present the mathematical model of fingerprint-based localization system and formulate the localization problem in Section~\ref{sect:prob}. The deep learning based localization approach is explained in Section~\ref{sec:DNN} and the corresponding performance is verified in Section~\ref{sect:results}. Finally, we conclude this paper in Section~\ref{sect:conc}.
\begin{figure}[t]
\centering
\subfigure[The experimental testbed]{
  \includegraphics[width=1.4 in]{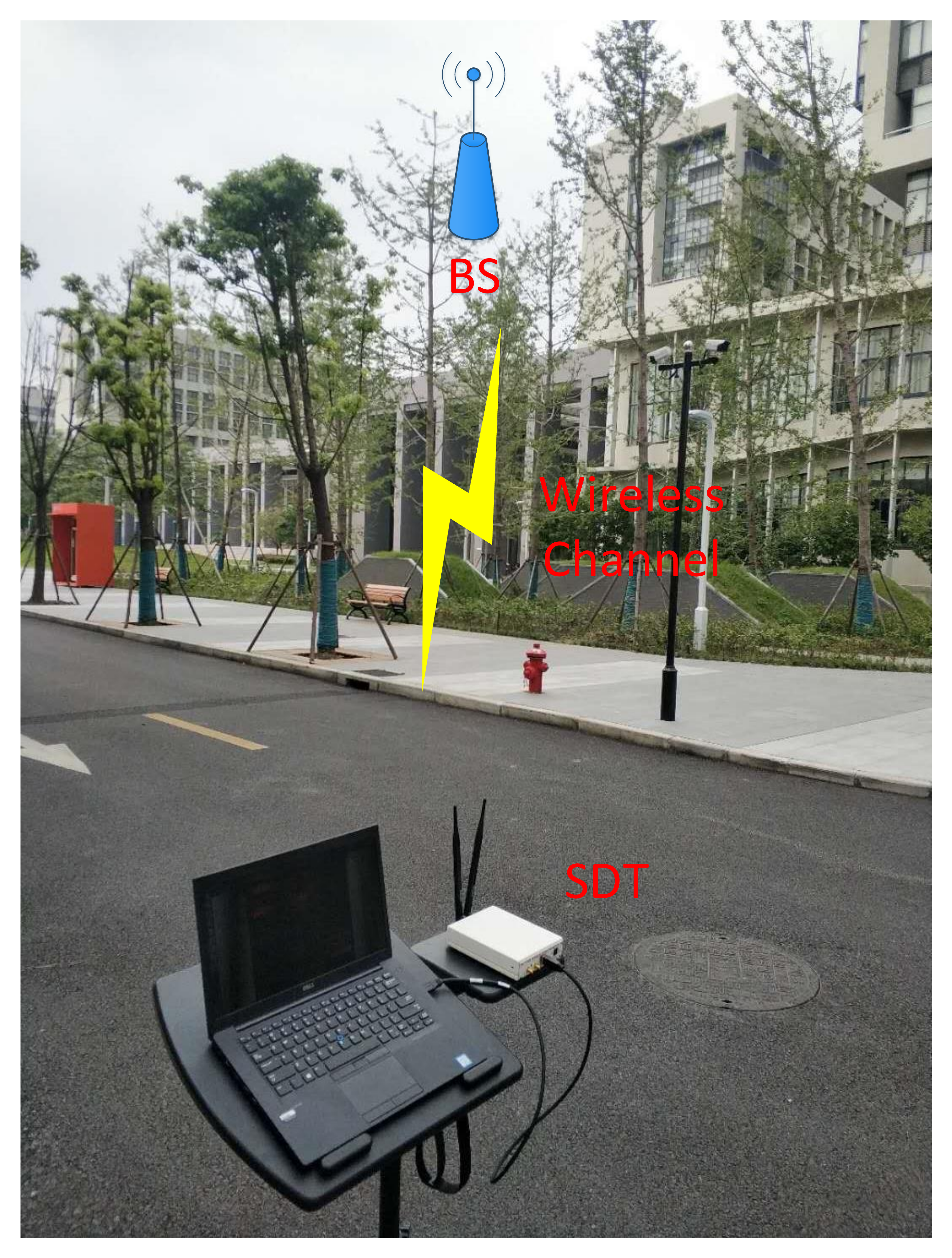}}
\hfill
\centering
\subfigure[System Architecture]{
  \includegraphics[width=1.6 in]{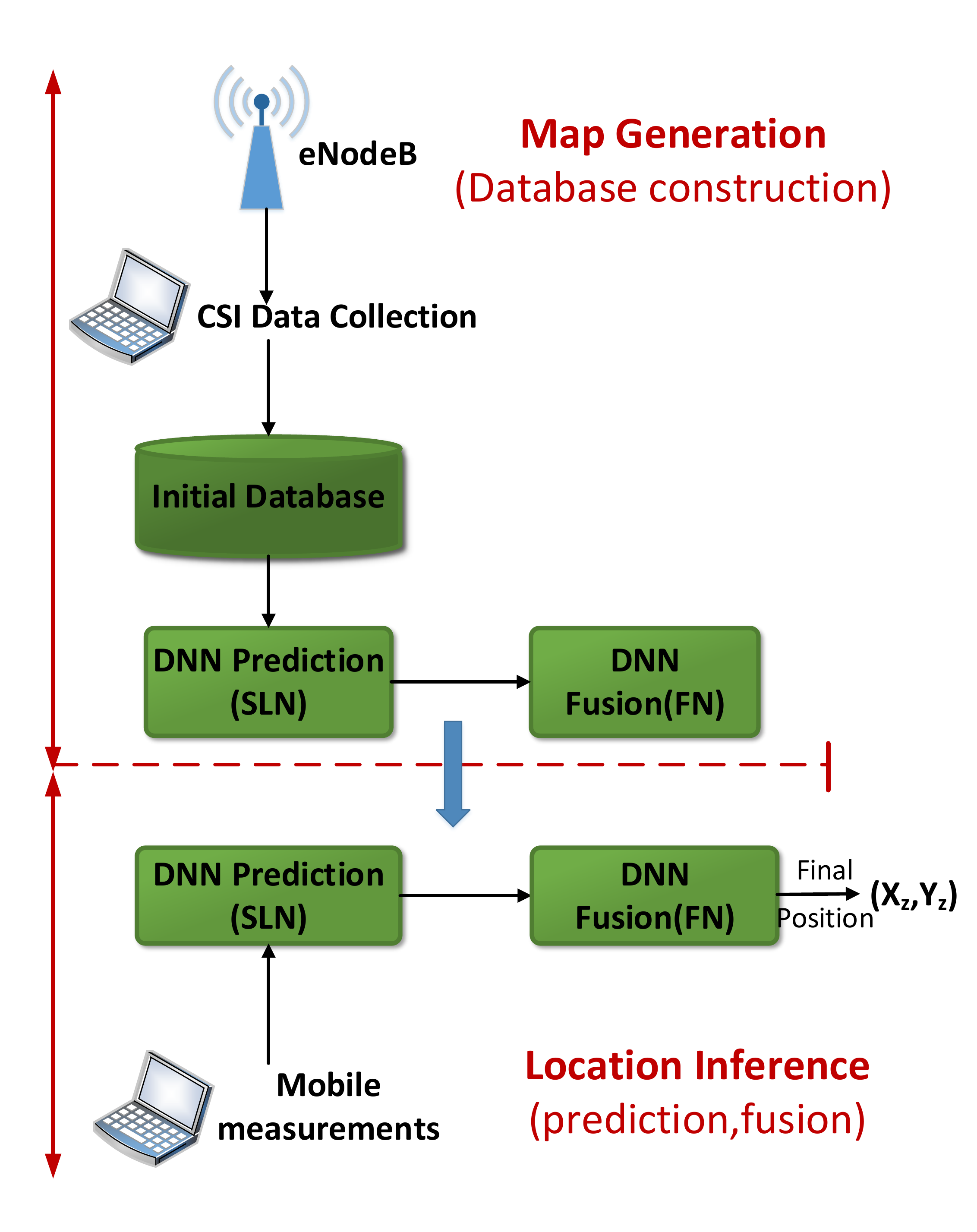}}
\caption{The proposed system configuration. (a) The experimental testbed in which SDT receives signals from BS and extracts location fingerprints. (b) The system architecture based on deep learning and LTE CSI fingerprint.}
\label{fig:system}
\end{figure}
\section{System Model} \label{sect:sys}

\subsection{System Configuration}
We consider a downlink LTE frequency division duplexing (FDD) system as shown in Fig.~\ref{fig:system} (a), where a software-defined terminal (SDT) is connected to the commercial LTE base station. The SDT, consisting of a laptop and a universal software-defined radio peripheral, performs the standard receiving process defined in 3GPP LTE Release 10\cite{3GPP:36.211}. Based on measuring reference signals, the instantaneous CSI can be obtained through the conventional channel estimation procedures.

The fingerprint-based localization can be roughly divided into two phases, the map generation phase and the location inference phase as shown in Fig.~\ref{fig:system} (b). In the map generation phase, the SDT collects different CSI values at different Reference Points (RP) to establish the initial CSI database and the deep learning algorithm based on DNN architecture is applied to extract the underlying features. Through this approach, we can obtain the fingerprint map using the extracted features and their corresponding RPs. In the location inference phase, the SDT collects real time CSIs and passes through this offline trained DNN to predict the user's location.
\begin{figure}
\centering
\includegraphics[width = 2.8 in]{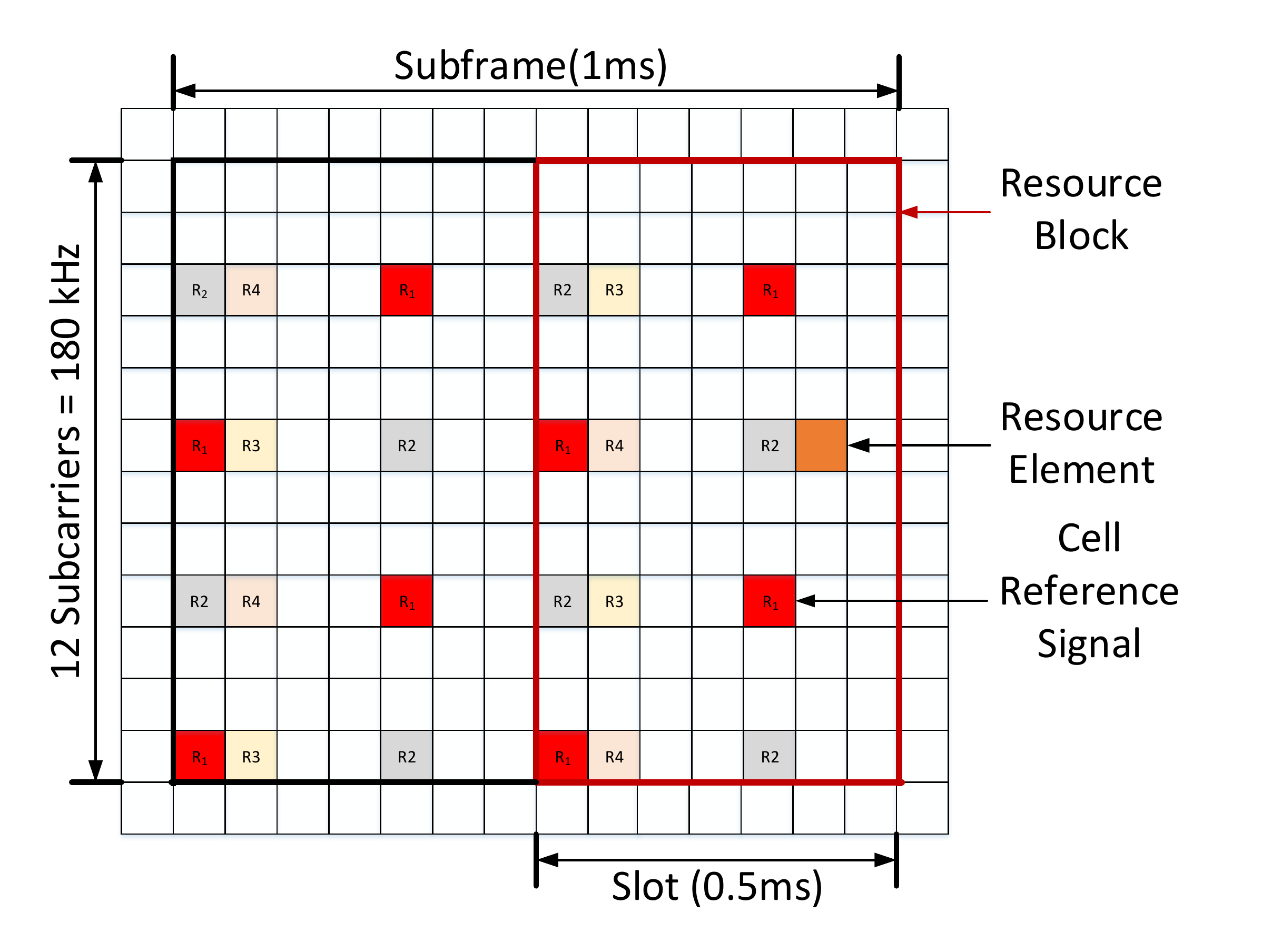}
\caption{The resource elements of Cell Specific Reference Signal for each transmitting antenna port in one subframe and one RB. }
\label{fig:resource_grid}
\end{figure}
\subsection{CSI Models}
\label{sub:csi}
Without loss of generality, we consider a resource element (RE) layout in the typical LTE subframe as shown in Fig.~\ref{fig:resource_grid}, where each (RE) corresponds to a subcarrier spacing of 15KHz in the frequency domain and an OFDM symbol duration of 66.7$\mu s$ in the time domain\cite{3GPP:36.211}. Following the standard LTE terminology, a resource block (RB) spans 12 consecutive subcarriers over a slot duration of 0.5 ms and the CSI knowledge is only estimated in the RE where cell reference signal (CRS)\footnote{CRS is a public unencrypted signal that can be used by all registered users.} is available.

Given the location $\mathcal{L}$ and the time slot $t$, the CSI knowledge from the $n^{th}$ transmit antenna at the $i^{th}$ subcarrier, $h_{i}^{n}(\mathcal{L},t) \in \mathbb{C}$, is given by,
\begin{equation}
h_{i}^{n}(\mathcal{L},t)=|h_{i}^{n}(\mathcal{L},t)|e^{j\cdot \theta_{i}^{n}(\mathcal{L},t)},
\end{equation}
where $|h_{i}^{n}(\mathcal{L},t)|$ and $\theta_{i}^{n}(\mathcal{L},t)$ denote the amplitude and phase information of the CSI $h_{i}^{n}(\mathcal{L},t)$, respectively, and $j$ stands for the imaginary unit. Since the phase information usually contains random jitters and noises due to the imperfect hardware components, and cannot be directly used without calibration \cite{shi2018accurate}, we rely on the amplitude information as the fingerprint information. For illustration purpose, we assemble all the estimated CSI at the location $\mathcal{L}$ and the time slot $t$ together, and the overall channel matrix, $\mathbf{H}(\mathcal{L}, t) \in \mathbb{R}^{N_{t} \times N_{c}}$, is given by,
\begin{flalign}
\begin{split}
&\mathbf{H}(\mathcal{L}, t)  = \\
&\left[
\begin{array}{c c c c}
|h_{CRS_{1}}^{1}(\mathcal{L},t)| & |h_{CRS_{1}}^{2}(\mathcal{L},t)| & \cdots & |h_{CRS_{1}}^{N_{t}}(\mathcal{L},t)| \\
|h_{CRS_{2}}^{1}(\mathcal{L},t)| & |h_{CRS_{2}}^{2}(\mathcal{L},t)| & \cdots & |h_{CRS_{2}}^{N_{t}}(\mathcal{L},t_{i})| \\
\cdots & \cdots & \cdots & \cdots \\
|h_{CRS_{N_{c}}}^{1}(\mathcal{L},t)| & |h_{CRS_{N_{c}}}^{2}(\mathcal{L},t)| & \cdots & |h_{CRS_{N_{c}}}^{N_{t}}(\mathcal{L},t)|
\end{array}
\right],
\end{split}&
\nonumber
\end{flalign}
where $N_{t}$ and $N_{c}$ denote the total number of transmit antennas and CRSs, and $CRS_{k}$ denotes the subcarrier index of the $k^{th}$ CRS.

The pathloss component, $\mathbf{H}(\mathcal{L})$, is obtained by averaging the instantaneous CSI magnitude, $\mathbf{H}(\mathcal{L},t)$, over sufficiently long transmission period.
Signal fingerprint-based localization techniques find the location of a device by comparing its signal pattern received from transmitters to a predefined database of signal patterns, e.g.,
\begin{eqnarray}
\hat{\mathcal{L}}_{m} = g \left(\{ \mathbf{H}(\mathcal{L}_{m})  \}\right),
\end{eqnarray}
where $\mathbf{H}(\mathcal{L}_{m})$ is the average fingerprint readings at location $m$. And the $\hat{\mathcal{L}}_{m}$ is estimated location according to mapping function $g(\cdot)$ and signal fingerprint $\mathbf{H}(\mathcal{L}_{m})$. Therefore,  the goal of localization system is to find the best mapping function that minimizes the difference between the estimated location and measured location.

\begin{figure}
\centering
\includegraphics[width=2.8 in]{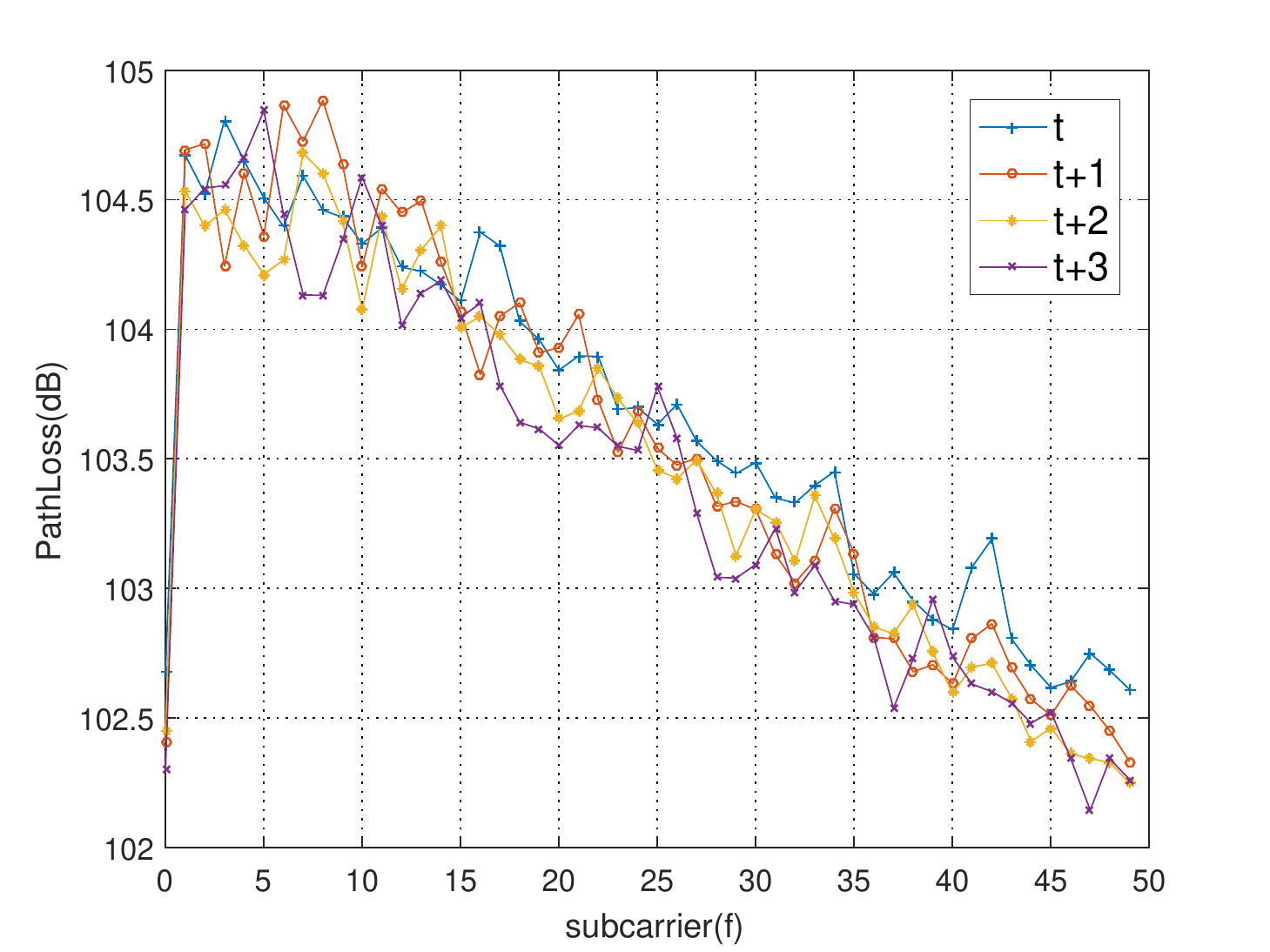}
\caption{CSI fluctuations over four consecutive time slots}
\label{fig:path_loss}
\end{figure}

However, due to the time-varying nature of CSI knowledge in the practical environment and the potential fluctuations of CSI amplitudes as shown in Fig.~\ref{fig:path_loss}, relying on the averaged value may lose some important information that can help to improve the localization accuracy. Therefore, in this paper, we collect the multiple CSI samples and estimate the location by identifying relation between them. It's worth noting that the dimensionality of CSI can be very high in the case of cellular networks with large bandwidth and multiple antennas. So it is difficult to directly find the relationship between multiple CSI samples. Inspired by data fusion of \cite{fang2011dynamic}, we map the relation between multiple CSI samples to the relation between multiple positions estimated by single CSI sample snapshot, which can be understood as dimension reduction and feature extraction. Then a time domain fusion approach is used to assemble multiple positioning estimations. The whole estimation process can be expressed as
\begin{eqnarray}
&& \qquad \  \hat{\mathcal{L}}_{m}^{{t}_{i}} = g_1 \left(\mathbf{H}(\mathcal{L}_{m},{t}_{i}) \right), \label{eqn:g1}\\
&&\hat{\mathcal{L}}_{m} = g_2 \left(\{\hat{\mathcal{L}}_{m}^{{t}_{i}}, t = 1, 2, \cdots, s\}  \right),\label{eqn:g2}
\end{eqnarray}
where $\hat{\mathcal{L}}_{m}^{t_{i}}$ is the estimated position according to fingerprint snapshot $\mathbf{H}(\mathcal{L}_{m},{t}_{i})$ and mapping function $g_1(\cdot)$. $\hat{\mathcal{L}}_{m}$ is the final estimated position according to multiple positions $\{\hat{\mathcal{L}}_{m}^{{t}_{i}}, t = 1, 2, \cdots, s\}$ and fusion method $g_2(\cdot)$.

\section{Problem Formulation}
\label{sect:prob}
\begin{figure*}[htbp]
\begin{center}
\includegraphics[width = 6 in]{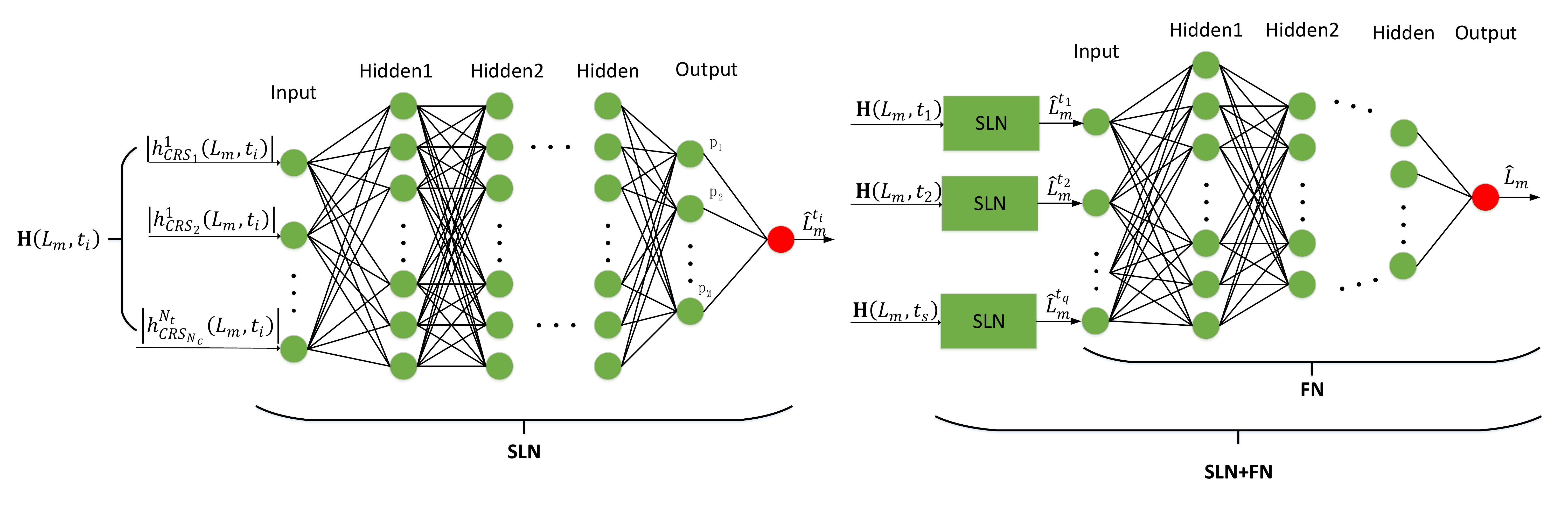}
\caption{The architecture of proposed deep neural networks, which consists of SLN and FN.}
\label{fig:dnn}
\end{center}
\end{figure*}

We divide the whole localization process into two steps: localization based single CSI fingerprint snapshot and multiple positions fusion. In the localization process, MDE is often used to describe the localization accuracy, where the corresponding mathematical expression, after averaging over $M$ localization periods, can be written as $\frac{1}{M} \sum_{m=1}^{M} \|\hat{\mathcal{L}}_m - \mathcal{L}_m\|$. Hence, we can model the localization process as two problems: slot-based localization and time domain fusion by using MDE minimization as follows
\begin{Prob}[Slot-based Localization]
\begin{eqnarray}
\underset{g_1 (\cdot)}{\textrm{min}} && \frac{1}{sM} \sum_{m=1}^{M}\sum_{i=1}^{s} \|\hat{\mathcal{L}}_m^{{t}_{i}} - \mathcal{L}_m\| \\
\textrm{s. t.} && (\ref{eqn:g1}), \nonumber\\
&& \hat{\mathcal{L}}_{m}^{{t}_{i}}, \mathcal{L}_{m} \in \mathcal{A}, \forall m, \forall i,
\end{eqnarray}
\end{Prob}
\begin{Prob}[Time Domain Fusion]
\begin{eqnarray}
\underset{g_{2} (\cdot)}{\textrm{min}} && \frac{1}{M} \sum_{m=1}^{M} \|\hat{\mathcal{L}}_m - \mathcal{L}_m\| \\
\textrm{s. t.} && (\ref{eqn:g2}), \nonumber \\
&& \hat{\mathcal{L}}_{m}^{{t}_{i}},\hat{\mathcal{L}}_{m}, \mathcal{L}_{m} \in \mathcal{A}, \forall i,m.
\end{eqnarray}
\end{Prob}
where $\mathcal{A}$ denotes the feasible localization areas.

Obviously, the above minimization is over all the possible functions of $g_1(\cdot)$ and $g_2(\cdot)$, so the solution is generally difficult to obtain.
The functions $g_1(\cdot)$ and $g_2(\cdot)$ have many heuristic proposals, such as KNN \cite{Pecoraro2018CSI} and WKNN \cite{Pecoraro2018CSI} for $g_1(\cdot)$ and MUltiple Classifiers mUltiple Samples (MUCUS) \cite{guo2017localization} and sliding window aided mode-based (SWIM) \cite{guo2018indoor}  for $g_2(\cdot)$. However, different from traditional fingerprints (such as RSSI/RSRP), the CSI based fingerprint incurs high dimensional nonlinear signal processing for feature extraction, where the classical models can not be applied \cite{fang2017learning}.

\section{Deep Learning Solutions}
\label{sec:DNN}

In this section, we propose to use deep learning based schemes to model the slot-based localization function $g_1(\cdot)$ and the time domain fusion function $g_2(\cdot)$.  Therefore, we rely on two stage cascaded neural networks, namely slot-based localization network (SLN) and fusion network (FN), as shown in Fig. \ref{fig:dnn}, to generate the intrinsic relations between measured CSIs and locations in the field trial experiments, where the detailed illustrations are as follows\footnote{The network design corresponds to situation of single antenna and CSI spanning 25 RBs. And we set $s=50$ in the experiments }.
\begin{table} [h]
\centering
\caption{Neural Network Architecture for SLN and FN}
\label{tab:nna}
\footnotesize
\begin{tabular}{c c c}
\toprule
\textbf{Layers}&\textbf{SLN}&\textbf{FN}\\
\midrule
Input Layer & $1 \times 50$ & $50 \times 2$ \\
\midrule
Hidden Layer 1 & Dense 256 + Dropout 0.3 & Dense 100 \\
\midrule
Hidden Layer 2 & Dense 256 + Dropout 0.3 & Dense 64 \\
\midrule
Hidden Layer 3 & Dense 256 + Dropout 0.3 & Dense 48\\
\midrule
Hidden Layer 4 & Dense 256 + Dropout 0.3 & Dense 12\\
\midrule
Output Layer & $1 \times 2$  & $1 \times 2$\\
\midrule
Loss Function & Cross Entropy & MSE\\
\bottomrule
\end{tabular}
\end{table}
\begin{figure*}[htbp]
\centering
\subfigure[Indoor Scenario]{
\begin{minipage}[t]{0.5\linewidth}
\centering
\includegraphics[width=2.5in]{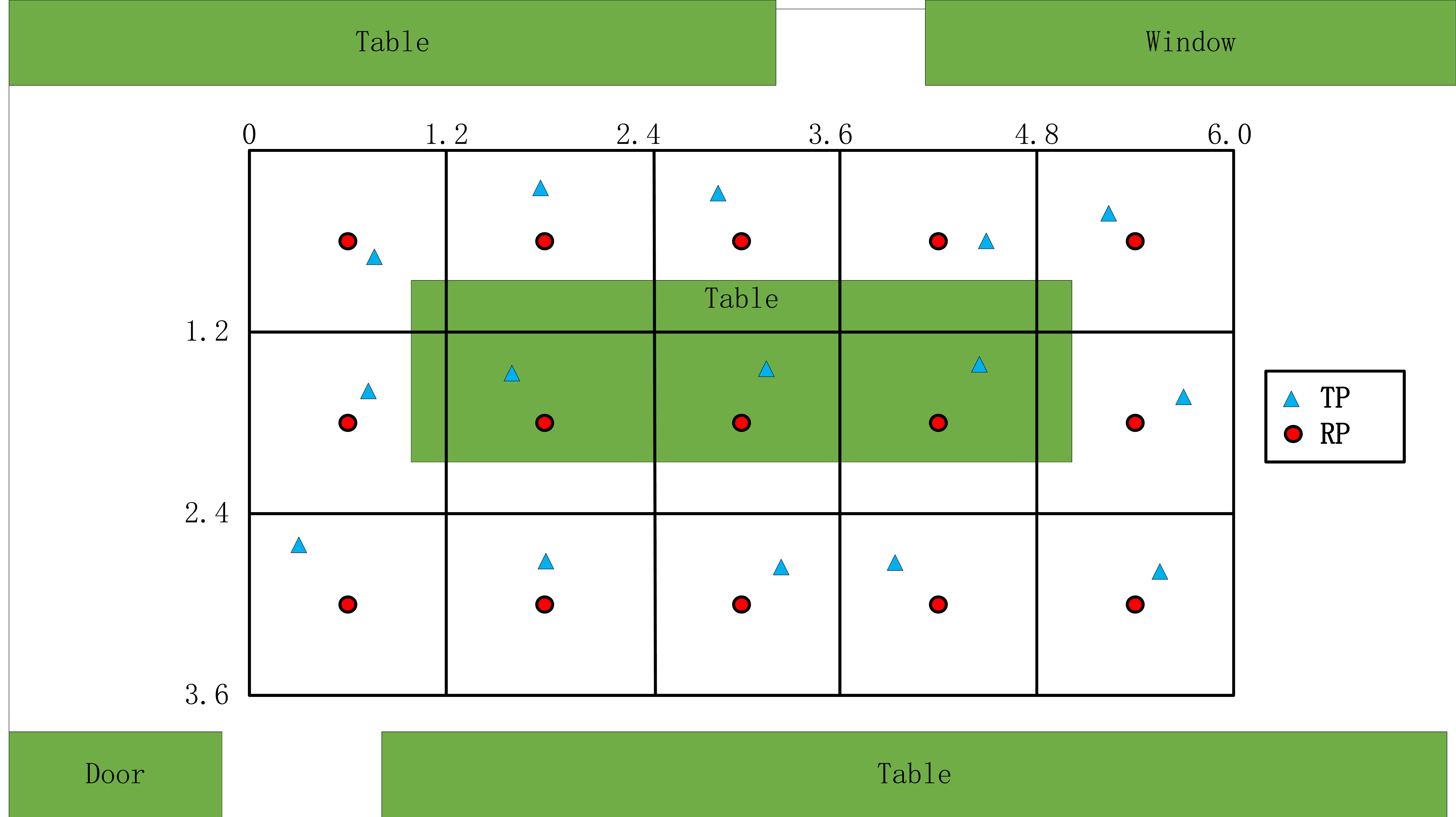}
\end{minipage}%
}%
\subfigure[Outdoor Scenario]{
\begin{minipage}[t]{0.5\linewidth}
\centering
\includegraphics[width= 2.7in]{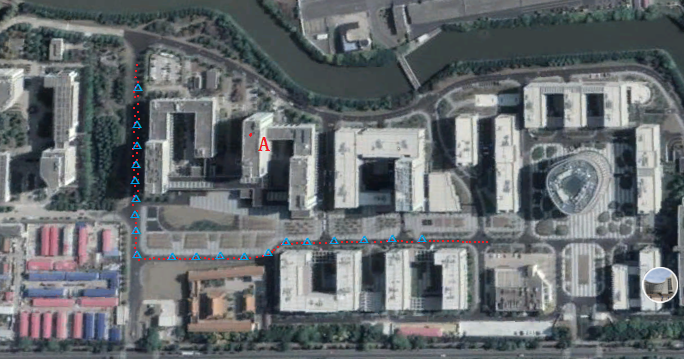}
\end{minipage}%
}%
\centering
\caption{The experimental areas of outdoor and indoor experiments, where the red points are RPs for establishing fingerprint database and blue triangles are the TPs for online testing. (a) Indoor environment. (b) Outdoor environment where point A is the position of commercial eNodeB.}
\label{fig:map}
\end{figure*}
\subsection{Neural Network Architecture}
To extract the features, we apply the multi-layer perceptron (MLP) methods \cite{seltzer2013investigation} for both SLN and FN to utilize the supervised learning. As shown in Fig. \ref{fig:dnn}, the input of SLN is single fingerprint snapshot $\mathbf{H}(\mathcal{L}_{m},{t}_{i})$. The output of SLN are the probabilities $\mathbf{p}=[p_1, p_2, \cdots, p_m]$ with respect to each RP and the general solution for localization is directly picking up the location with the largest probability. To improve the localization accuracy, we use weighting methods to estimate location, which can written as $\hat{\mathcal{L}}_{m}^{t_{i}} = \sum_{m=1}^{M}\hat{\mathcal{L}}_m\times p_{m}$. The input of FN is the multiple positions $\{\hat{\mathcal{L}}_{m}^{t_{i}}, t = 1, 2, \cdots, s \}$ estimated by SLN according to $s$  time-continuous fingerprint snapshots. In each hidden layer, we use the {\em rectified linear unit} (ReLU) \cite{nair2010rectified} as the activation function, and in the output layers,  {\em softmax} and {\em linear} are applied as the activation functions for SLN and FN, respectively. The loss functions for SLN and FN are {\em Cross-Entropy} (CE) and  {\em Mean Squared Error} (MSE), which can be written as $L_{SLN}=-\sum_{m=1}^{M}p_{m} \times \log(q_{m})$ and $L_{FN}=\frac{1}{M}\sum_{m=1}^{M} \|\hat{\mathcal{L}}_m - \mathcal{L}_m\|^{2}$,
with $p_{m}$ and $q_{m}$ representing the probability distribution of ground truth $\mathcal{L}_m$ and estimated positions $\hat{\mathcal{L}}_{m}^{t}$ in the training stage. The detailed information for SLN and FN are summarized in Table~\ref{tab:nna}.

\subsection{Training Data Sets}
We pre-define $M$ reference points (RPs) in the target area and record the CSI variations at each RP for a period of time. Through this approach, we have $M$ locations $\{\mathcal{L}_m\}$ and the corresponding measured CSIs $\{\mathbf{H}(\mathcal{L}_m,t)\}$, which can be used as supervised information to train the SLN. Once the training of SLN is finalized, we can use the trained network to model $g_1(\cdot)$ and obtain $\hat{\mathcal{L}}_{m}^{t_{i}}$ for the $m^{th}$ period according to \eqref{eqn:g1}. 
Together with $M$ locations $\{\mathcal{L}_m\}$, we have the training data sets for FN.

\section{Experimental Results} \label{sect:results}

In this section, we provide some numerical examples to verify the effectiveness of proposed localization methods by comparing with existing fingerprint based localization scheme in the indoor and outdoor environments. In the indoor scenario, a 3.6m $\times$ 6m rectangular area containing 15 RPs is covered, and in the outdoor scenario, the field trial measurements are performed in the campus of {\em Shanghai University} with a roughly 360m $\times$ 195m suburban area with 105 RPs. The RPs for the fingerprint map generation are uniformly distributed with interval 1.2m in the indoor case and along the moving trajectory with interval 5m in outdoor case, respectively, and the testing points (TPs) are randomly selected as shown in Fig. \ref{fig:map}. The CSI is measured from a commercial base station served by the operator {\em China Telecom}. The cumulative distribution function (CDF) of the localization errors is used to evaluate the performance of the proposed fingerprint localization system.

\subsection{Indoor Environment}

\begin{figure}
\centering
\includegraphics[width = 2.8 in]{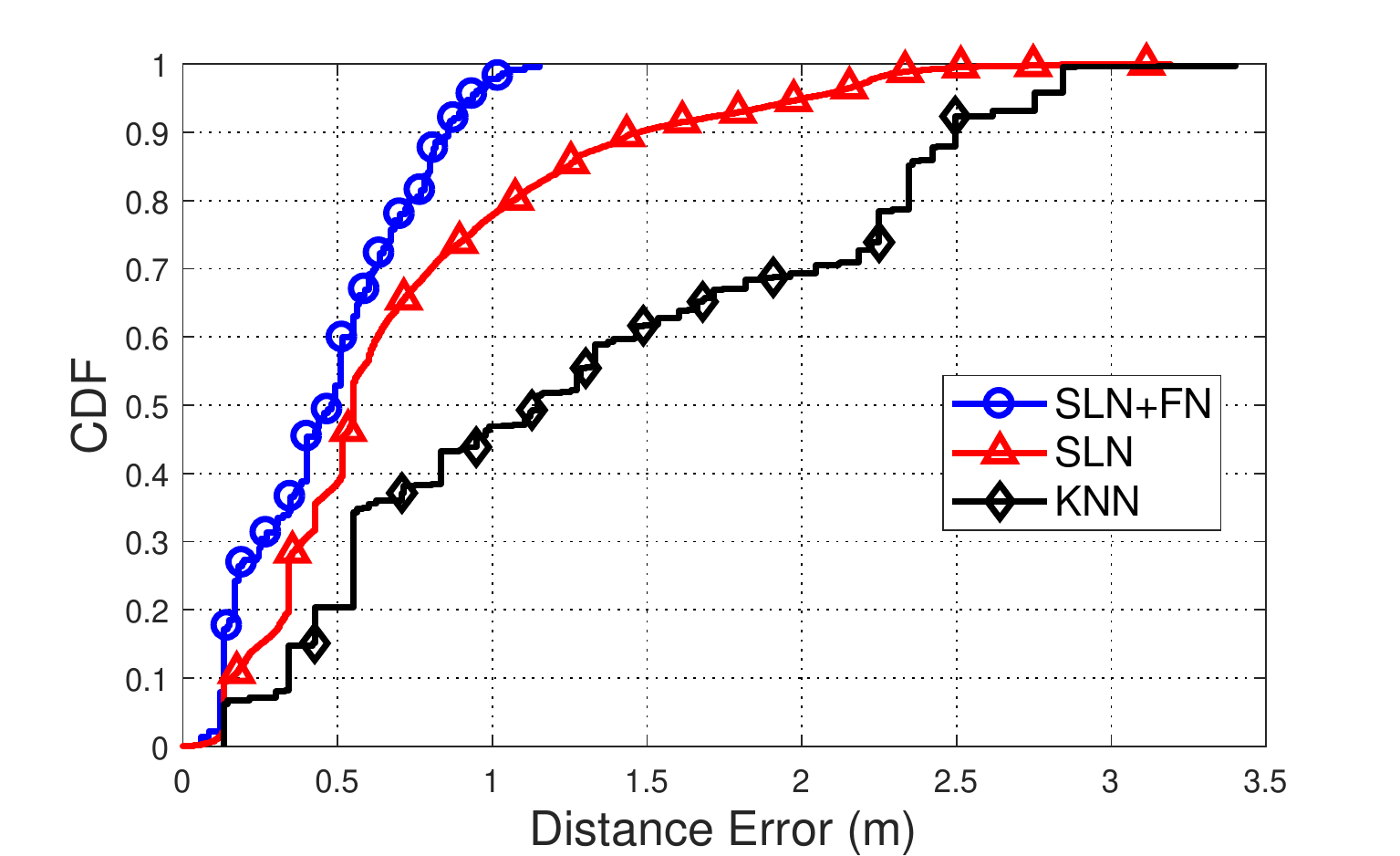}
\caption{CDF of localization error in indoor environment. }
\label{fig:indoor_result}
\end{figure}

 Fig.~\ref{fig:indoor_result} shows that the SLN based on fingerprint snapshot and deep learning has a certain improvement in accuracy compared with the traditional method KNN in indoor scenario. But they both have a poor robustness in terms of maximum distance error, which is caused by fluctuation of CSI fingerprint. The accuracy and robustness of system both have a significant improvement by combining SLN and time fusion algorithm FN. More precisely, the MDE and the maximum distance error for the proposed localization scheme SLN+FN can be reduced to 0.47m and 1.15m, respectively, which is nearly three times better than the traditional KNN approach.

\subsection{Outdoor Environment}
\begin{figure}
\centering
\includegraphics[width = 2.8 in]{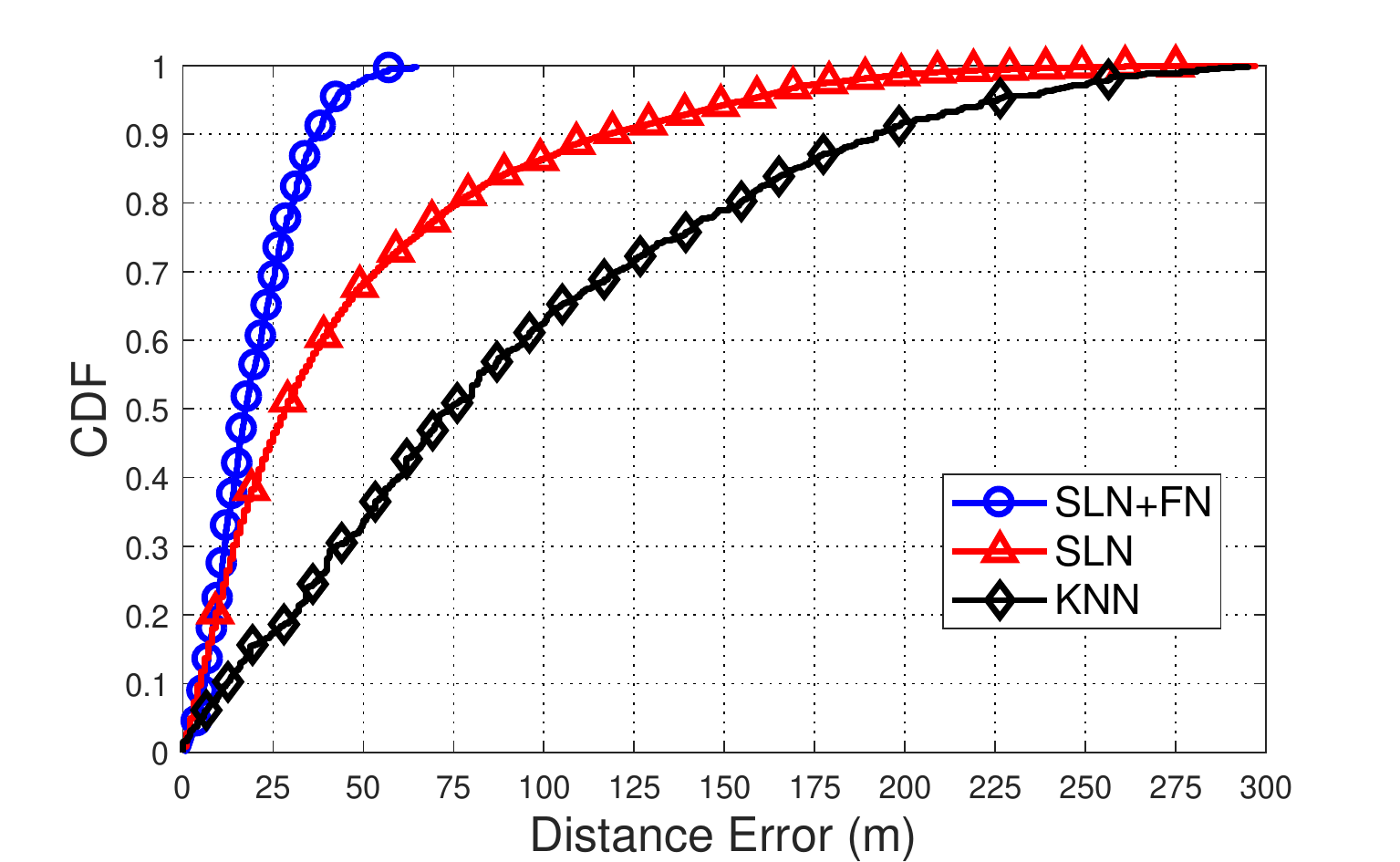}
\caption{CDF of localization error in outdoor environment. }
\label{fig:outdoor_result}
\end{figure}

The CDF of distance errors for different localization schemes in the outdoor environment is provided in Fig.~\ref{fig:outdoor_result}, where the absolute values of the MDE and the maximum distance error are much higher due to  channel fluctuation and larger coverage area. However, the proposed localization scheme SLN+FN provides more than four times better in the MDE performance, e.g. from 90.9m to 19.9m, and in the maximum distance error performance, e.g. from 295m to 64.8m, compared with KNN. In addition, if compared with the SLN, the SLN+FN also offers more accurate and robust localization results.




\section{Conclusion} \label{sect:conc}
In this paper, we utilize field trial experiments to investigate the feasibility of fingerprint based localization using real time measured CSI knowledge from commercial LTE base stations. In particular, a two stage deep learning based method is proposed to extract the intrinsic features and conduct time domain fusion. Preliminary results demonstrate that the proposed localization technique can achieve the MDE of 0.47m and 19.9m, and the maximum distance error of 1.15m and 64.8m in indoor and outdoor environments, respectively.

\section*{Acknowledgement}
This work was supported by the National Natural Science Foundation of China (NSFC) Grants under No. 61701293 and No. 61871262, the National Science and Technology Major Project Grants under No. 2018ZX03001009, the Huawei Innovation Research Program (HIRP), and research funds from Shanghai Institute for Advanced Communication and Data Science (SICS).

\bibliographystyle{IEEEtran}
\bibliography{IEEEabrv,rf}

\begin{thebibliography}{10}
\providecommand{\url}[1]{#1}
\csname url@samestyle\endcsname
\providecommand{\newblock}{\relax}
\providecommand{\bibinfo}[2]{#2}
\providecommand{\BIBentrySTDinterwordspacing}{\spaceskip=0pt\relax}
\providecommand{\BIBentryALTinterwordstretchfactor}{4}
\providecommand{\BIBentryALTinterwordspacing}{\spaceskip=\fontdimen2\font plus
\BIBentryALTinterwordstretchfactor\fontdimen3\font minus
  \fontdimen4\font\relax}
\providecommand{\BIBforeignlanguage}[2]{{%
\expandafter\ifx\csname l@#1\endcsname\relax
\typeout{** WARNING: IEEEtran.bst: No hyphenation pattern has been}%
\typeout{** loaded for the language `#1'. Using the pattern for}%
\typeout{** the default language instead.}%
\else
\language=\csname l@#1\endcsname
\fi
#2}}
\providecommand{\BIBdecl}{\relax}
\BIBdecl

\bibitem{del2017survey}
J.~A. del Peral-Rosado, R.~Raulefs, J.~A. L{\'o}pez-Salcedo, and
  G.~Seco-Granados, ``Survey of cellular mobile radio localization methods:
  from 1{G} to 5{G},'' \emph{IEEE Communications Surveys \& Tutorials},
  vol.~20, no.~2, pp. 1124--1148, Dec. 2018.

\bibitem{geiger2012we}
A.~Geiger, P.~Lenz, and R.~Urtasun, ``Are we ready for autonomous driving? the
  kitti vision benchmark suite,'' in \emph{IEEE Conf. on Computer Vision and
  Pattern Recognition (CVPR)}, Jun. 2012.

\bibitem{Pecoraro2018CSI}
G.~Pecoraro, S.~D. Domenico, E.~Cianca, and M.~D. Sanctis, ``C{SI}-based
  fingerprinting for indoor localization using lte signals,'' \emph{Eurasip
  Journal on Advances in Signal Processing}, vol. 2018, no.~1, p.~49, 2018.

\bibitem{shi2018accurate}
S.~Shi, S.~Sigg, L.~Chen, and Y.~Ji, ``Accurate location tracking from
  csi-based passive device-free probabilistic fingerprinting,'' \emph{IEEE
  Transactions on Vehicular Technology}, vol.~67, no.~6, pp. 5217--5230, Feb.
  2018.

\bibitem{ye2017neural}
X.~Ye, X.~Yin, X.~Cai, A.~P. Yuste, and H.~Xu, ``Neural-network-assisted ue
  localization using radio-channel fingerprints in lte networks,'' \emph{IEEE
  Access}, vol.~5, pp. 12\,071--12\,087, Jun. 2017.

\bibitem{r9}
\BIBentryALTinterwordspacing
{3GPP Release} 9. [Online]. Available:
  \url{http://www.3gpp.org/specifications/releases/71-release-9}
\BIBentrySTDinterwordspacing

\bibitem{wu2017mitigating}
C.~Wu, Z.~Yang, Z.~Zhou, Y.~Liu, and M.~Liu, ``Mitigating large errors in
  wifi-based indoor localization for smartphones,'' \emph{IEEE Transactions on
  Vehicular Technology}, vol.~66, no.~7, pp. 6246--6257, Nov. 2017.

\bibitem{wang2017csi}
X.~Wang, L.~Gao, S.~Mao, and S.~Pandey, ``C{SI}-based fingerprinting for indoor
  localization: A deep learning approach,'' \emph{IEEE Transactions on
  Vehicular Technology}, vol.~66, no.~1, pp. 763--776, Mar. 2017.

\bibitem{machaj2011rank}
J.~Machaj, P.~Brida, and R.~Pich{\'e}, ``Rank based fingerprinting algorithm
  for indoor positioning,'' in \emph{IEEE Indoor Positioning and Indoor
  Navigation (IPIN)}, 2011, pp. 1--6.

\bibitem{lin2014group}
T.-N. Lin, S.-H. Fang, W.-H. Tseng, C.-W. Lee, and J.-W. Hsieh, ``A
  group-discrimination-based access point selection for wlan fingerprinting
  localization,'' \emph{IEEE Transactions on Vehicular Technology}, vol.~63,
  no.~8, pp. 3967--3976, Jan. 2014.

\bibitem{3GPP:36.211}
\emph{3rd Generation Partnership Project; Technical Specification Group Radio
  Access Network;Evolved Universal Terrestrial Radio Access (E-UTRA); Physical
  channels and modulation (Release 10)}, 3GPP, Dec. 2012, {TS} 36.211.

\bibitem{fang2011dynamic}
S.-H. Fang, Y.-T. Hsu, and W.-H. Kuo, ``Dynamic fingerprinting combination for
  improved mobile localization,'' \emph{IEEE Transactions on Wireless
  Communications}, vol.~10, no.~12, pp. 4018--4022, 2011.

\bibitem{guo2017localization}
X.~Guo and N.~Ansari, ``Localization by fusing a group of fingerprints via
  multiple antennas in indoor environment,'' \emph{IEEE Transactions on
  Vehicular Technology}, vol.~66, no.~11, pp. 9904--9915, Jul. 2017.

\bibitem{guo2018indoor}
X.~Guo, N.~Ansari, L.~Li, and H.~Li, ``Indoor localization by fusing a group of
  fingerprints based on random forests,'' \emph{IEEE Internet of Things
  Journal}, 2018.

\bibitem{fang2017learning}
S.-H. Fang, Y.-X. Fei, Z.~Xu, and Y.~Tsao, ``Learning transportation modes from
  smartphone sensors based on deep neural network,'' \emph{IEEE Sensors
  Journal}, vol.~17, no.~18, pp. 6111--6118, Aug. 2017.

\bibitem{seltzer2013investigation}
M.~L. Seltzer, D.~Yu, and Y.~Wang, ``An investigation of deep neural networks
  for noise robust speech recognition,'' in \emph{Proc. ICASSP}, 2013.

\bibitem{nair2010rectified}
V.~Nair and G.~E. Hinton, ``Rectified linear units improve restricted boltzmann
  machines,'' in \emph{Proc. Int'l Conf. Machine Learning}, 2010.

\end{thebibliography}

\end{document}